\documentstyle[prd,aps,epsf]{revtex}
\begin{document}

\thispagestyle{plain}
{
\setlength{\topskip}{0pt}
\baselineskip=0pt
\leftline{\baselineskip8pt\vbox to0pt{\hbox{}\hbox{}\vss}}
\rightline{\baselineskip8pt\vbox to10pt{\hbox{KUNS-1679}
                                                  \hbox{YITP-00-46}\vss}}
}
\begin{center}
{\large \bf
Instability of a membrane intersecting a black hole
}
\end{center}
\begin{center}
Susumu Higaki\footnote{email:higaki@tap.scphys.kyoto-u.ac.jp}${}^{1}$,
Akihiro Ishibashi\footnote{email:akihiro@yukawa.kyoto-u.ac.jp}${}^{2}$ and
Daisuke Ida\footnote{email:ida@tap.scphys.kyoto-u.ac.jp}${}^{1}$
\\
\vspace{1em}
${}^{1}${\it Department of Physics, Kyoto University, Kyoto 606-8502, Japan}
\\
${}^{2}${\it Yukawa Institute for Theoretical Physics, Kyoto University,
Kyoto 606-8502, Japan}
\end{center}

\begin{abstract}
The stability of a Nambu-Goto membrane at the equatorial plane of the
Reissner-Nordstr{\o}m-de Sitter spacetime is studied.
The covariant perturbation formalism is applied to study the behavior of
the perturbation of the membrane.
The perturbation equation is solved numerically.
It is shown that a membrane intersecting a charged black hole, including
extremely charged one, is unstable and that the positive cosmological
constant
strengthens the instability.
\end{abstract}
\draft

\section{INTRODUCTION}
It is believed that in the early universe a series of vacuum phase
transitions led to several types of topological defects~\cite{KTVS}.
Topological defects are relics of the early universe and are expected to
convey some information on physics of very high energy scales beyond our
reach
with ground-based accelerators.
On the other hand, topological defects are candidates for the seed of the
observed large scale structure of the universe such as sheet-like or
filamentary structures or voids.
Thus topological defects are attractive examples which connect the
high energy physics and cosmology and we might be able to have some
information on high energy physics through cosmological observations such as
gravitational wave detection in future.
Topological defects, if they existed, would interact with other strong
gravitational sources such as black holes, and then they might have
experienced large deformation and emitted some information on themselves as
gravitational waves.
If we succeed in detecting such gravitational waves and identifying them,
we will be able to confirm the existence of topological defects and hence
the occurrence of a vacuum phase transition.

The topological defects with finite extent such as the cosmic string or the
domain wall are known to produce an unusual gravitational field~\cite{Ip,Vi}
and their dynamics are slightly complicated.
There are some studies on the interaction between a cosmic string
and a black hole with the hope of detecting gravitational waves from cosmic
strings.
De Villier and Frolov studied~\cite{DF1,DF2} the dynamics of the scattering
and capturing process of an infinitely thin test string by a Schwarzschild
black hole.
However less is known on the interaction between a domain wall and a black
hole; only the static configurations have been studied so far.
Morisawa et al.~\cite{Mo} showed the existence of static configurations
of a thick domain wall intersecting a Schwarzschild black hole in its
equatorial plane.
Christensen et al.~\cite{Chr} numerically found a series of static
configurations of a domain wall as a Nambu-Goto membrane in the
Schwarzschild
spacetime, some of which represent an intersecting pair of a domain wall
and a black hole.
When arranged in sequence, these configurations seem to represent a
scattering
and capturing process of a domain wall by a black hole.
In most situations of cosmological interest, the thickness of a domain wall
will be much smaller than the horizon radius of a solar-mass or primordial
black hole, so that the infinitely thin wall (membrane) approximation will
be
valid in such cases.
When we further neglect the gravitational effect of a domain wall, its
spacetime history, i.e., the world sheet is governed by the Nambu-Goto
action
and it is called the Nambu-Goto membrane.

A membrane lying in the equatorial plane of the black hole spacetime has
the highest symmetry among the configurations which represent an
intersecting
pair of a membrane and a black hole.
This simple configuration is a possible candidate for a final state of the
scattering and capturing process provided such a state stably exists.
Having this in mind, we study the stability of a Nambu-Goto membrane at the
equatorial plane of the Reissner-Nordstr{\o}m-de Sitter (RNdS) spacetime.
The charge of the black hole and the positive cosmological constant are
included to see their effect on the stability.
This mimics the situation where a domain wall resulting from a vacuum phase
transition is intersecting a charged primordial black hole in the
early universe.
We consider the linear perturbation of the membrane by means of the
covariant
perturbation formalism~\cite{GV,Gu}.
The perturbation equation reduces to a wave equation on the world
sheet with the mass term, which is negative in the presence of the charge of
the black hole or the positive cosmological constant.
Accordingly instability is naively expected.
Solving the perturbation equation reduces to solving a two point boundary
value problem.
We numerically solve it by the shooting and the relaxation methods.
We find that a membrane intersecting a RNdS black hole is unstable.
We find the instability at least in the range of $0.001\le e/M \le 1$ for
$\Lambda >0$, and $0.75\le e/M \le 1$ for $\Lambda =0$, whereas the membrane
in the Schwarzschild background is stable.

While our treatment of a domain wall is simple with both its thickness and
gravitational effect neglected, our model of the membrane-black hole system
is all the more elementary and relevant in other contexts.
The interaction between black holes and extended objects is important as
an elementary physical process.
Indeed much attention has recently been paid to the interaction between
black
holes and extended objects such as membranes or strings in the
string/M-theory~\cite{Gre,GL1,GL2,GL3}, the brane world
scenario~\cite{Cham,Em,GS,RS1,RS2} etc..
For instance in the context of the brane world models, our universe is
described as a four-dimensional domain wall in the five dimensional bulk
spacetime.
Some people argue that a black hole on the gravitating membrane is realized
as a ``black cigar'' in the bulk spacetime which intersects the
membrane~\cite{Cham}.

The rest of the paper is organized as follows.
In Sec.\ref{eqs} we review the covariant perturbation formalism and specify
the perturbation equation and the boundary conditions.
The numerical algorithm is also explained there.
In Sec.\ref{results} we present the results of numerical and analytical
considerations. Finally we summarize and discuss the results in
Sec.\ref{discussion}.

\section{BASIC EQUATIONS}\label{eqs}
\subsection{Equation of motion}
The history of a membrane (world sheet) is described by the timelike
hypersurface $(\Sigma, \gamma_{ab})$ embedded in the
four-dimensional spacetime $(M, g_{\mu\nu})$. The embedding is given by
\begin{equation}
x^{\mu}=X^{\mu}({\xi }^a ) \hspace{2em}
\mu =0,\cdots ,3, \hspace{1em} a=0,\cdots ,2,
\end{equation}
where $x^{\mu}$'s are the spacetime coordinates and ${\xi}^a $'s the world
sheet coordinates.
The induced metric
${\gamma}_{ab}$ on $\Sigma$ is given by
\begin{equation}
{\gamma}_{ab}=X^{\mu}_{,a} X^{\nu}_{,b} g_{\mu\nu}\label{indmetric}.
\end{equation}
The dynamics of the membrane is described by the Nambu-Goto
action
\begin{equation}
S[X^{\mu},X^{\mu}_{,a}]=-\sigma\int_{\Sigma }d^3\xi \sqrt{-\gamma}
                        \label{NGaction},
\end{equation}
where $\sigma $ represents the surface energy density of the membrane in
its rest frame.

We consider the variation of the action with respect to $X_{\mu}$,
\begin{equation}
\frac{\delta S}{\delta X_{\mu}}
=\sigma
({\Gamma}^{\mu}_{\alpha\beta}\gamma^{ab}X^{\alpha}_{,a}X^{\beta}_{,b}
+\Box X^{\mu}) \label{variation},
\end{equation}
where ${\Gamma}^{\mu}_{\alpha\beta}$ is the spacetime affine connection and
$\Box $ is the d'Alembertian on the world sheet
\begin{equation}
\Box =\frac{1}{\sqrt{-\gamma}}
\frac{\partial}{\partial {\xi}^{a}}(\sqrt{-\gamma}{\gamma }^{ab}
\frac{\partial}{\partial {\xi}^{b}}) .
\end{equation}
Recall the Gauss-Weingarten equation
\begin{equation}
D_b X^{\mu }_{,a} +{\Gamma }^{\mu }_{\alpha\beta }
X^{\alpha }_{,a}X^{\beta }_{,b} =K_{ab}n^{\mu } \label{gausswein},
\end{equation}
where $K_{ab}$ is the extrinsic curvature defined as
\begin{equation}
K_{ab}=-X^{\mu }_{,a}X^{\nu }_{,b}{\nabla }_{\nu }n_{\mu },
\end{equation}
$n_{\mu }$ is the normal to the world sheet, $D_a$ is the world sheet
covariant
derivative and ${\nabla }_{\mu }$ is the spacetime covariant derivative.
Contracting Eq.~(\ref{gausswein}) with ${\gamma }^{ab}$, we obtain
\begin{equation}
\Box X^{\mu }+{\Gamma }^{\mu }_{\alpha\beta }{\gamma }^{ab}
X^{\alpha }_{,a} X^{\beta }_{,b}={\gamma }^{ab}K_{ab}n^{\mu } .
\end{equation}
With this equation Eq.~(\ref{variation}) becomes
\begin{equation}
\frac{\delta S}{\delta X_{\mu }}=\sigma {\gamma }^{ab}K_{ab}
n^{\mu}=0 \label{variation2}.
\end{equation}
Eq.~(\ref{variation2}) has only the component perpendicular to the world
sheet.
Hence the variation parallel to the world sheet has no physical meaning.
Finally the equation of motion of the membrane is
\begin{equation}
K={\gamma }^{ab}K_{ab}=0\label{eom}.
\end{equation}
This is the equation of minimal surfaces.
In general, Eq.~(\ref{eom}) cannot be solved analytically.

\subsection{Perturbation equation}
As noted above, the physically meaningful measure of the perturbation
$\delta X^{\mu }$ is the scalar
\begin{equation}
\Phi =n_{\mu }\delta X^{\mu }.
\end{equation}
Taking the variation of Eq.~(\ref{eom}) with respect to $\Phi$ and setting
it to zero, we obtain the perturbation equation
\begin{equation}
\Box\Phi -\left( {}^{3}R-{}^{4}R_{\mu\nu}h^{\mu\nu}\right)\Phi=0
\label{perturb},
\end{equation}
where $h_{\mu\nu}\equiv g_{\mu\nu}-n_{\mu}n_{\nu}$ is the projection operator
onto
the world sheet.

\subsection{Perturbation of a membrane at the equatorial plane of the RNdS
spacetime}
A membrane lying in the equatorial plane of the black hole spacetime has
the highest symmetry among the configurations found by~\cite{Chr} which
represent an intersecting pair of a membrane and a Schwarzschild black hole.
This simple configuration is a possible candidate for a final state of the
scattering and capturing process of a membrane and a black hole provided
such a state stably exists.

In general, spherically symmetric black hole spacetimes have the discrete
symmetry about the equatorial plane, i.e., the equatorial plane consists of
fixed points of this symmetry.
Hence the equatorial plane is totally geodesic and $K=0$.

The most general spherically symmetric black hole spacetime is the
Reissner-Nordstr{\o}m-de Sitter spacetime
\begin{equation}
g_{\mu\nu}dx^{\mu}dx^{\nu}=-f d t^{2}+f^{-1}d r^{2}+r^{2}d{\theta}^{2}+
r^{2}{\sin}^{2}\theta d{\varphi}^{2} ,
\end{equation}
where
\begin{equation}
f = 1-\frac{2M}{r}+\frac{e^2}{r^2}-\frac{\Lambda }{3}r^2 \label{f},
\end{equation}
$\Lambda $ is the cosmological constant and $M$ and $e$ are the mass and
charge of the black hole, respectively.
The relation between the spacetime coordinates $\{t, r, \theta , \varphi \}$
and the world sheet coordinates $\{T, R, \Psi \}$ of the membrane at the
equatorial plane is simply
\begin{equation}
t=T,\hspace{1em} r=R,\hspace{1em} \theta =\frac{\pi}{2},\hspace{1em}
\varphi =\Psi .
\end{equation}
Then the induced metric (\ref{indmetric}) is
\begin{equation}
{\gamma}_{ab}d{\xi}^{a}d{\xi}^{b}=-f d T^{2}+f^{-1}d{R}^{2}+R^{2}d{\Psi}^{2}
.
\end{equation}
This means that the geometry of the membrane is a three-dimensional black
hole
spacetime.
The perturbation equation (\ref{perturb}) becomes
\begin{equation}
\Box\Phi +\left( \frac{e^2}{r^4}+\Lambda \right)\Phi =0 \label{perturb2}.
\end{equation}
The mass term becomes negative when the charge of the black
hole or the positive cosmological constant is present, which implies
instability.
This is why we consider the equatorial plane of the
Reissner-Nordstr{\o}m-de Sitter spacetime.
In addition, it mimics a cosmologically interesting situation where a domain
wall resulting from a vacuum phase transition is intersecting a charged
primordial black hole in the early universe.

By a separation of variables, $\Phi =(\chi (r)/\sqrt{r})
\exp (i\omega t +im\varphi )$
, Eq.~(\ref{perturb2}) is transformed into a Schr\"{o}dinger type equation
in
the three-dimensional black hole spacetime
\begin{equation}
-\frac{d^2 \chi}{dr_{*}^2}+V(r)\chi ={\omega }^2 \chi\label{Schroedinger} ,
\end{equation}
\begin{equation}
V(r)=f\left[\left(\frac{m^2}{r^2} -\frac{e^2}{r^4} -\Lambda\right) -
       \frac{f -2f'r}{4r^2}\right]\label{potential} ,
\end{equation}
where $r_{*}=\int^r dr/f$ is the usual tortoise coordinate.
For the cases $M\ne 0$ we normalize $r, e, \Lambda ,\omega$ as
\begin{equation}
r/M \to r, e/M \to e, \Lambda M^2 \to \Lambda , \omega M \to \omega .
\label{normalization}
\end{equation}

\subsection{Boundary conditions}\label{2.d}
The present problem reduces to solving a Schr\"{o}dinger type equation, and
so does the problem of the metric perturbation of a four-dimensional black
hole spacetime.
Hence we will follow the stability analysis of the latter.

We consider the cases $M\ne 0$.
Then the tortoise coordinate $r_{*}$ goes from $-\infty$ to $+\infty$.
$r_{*}=-\infty$ corresponds to the event horizon and $r_{*}=+\infty$ to
infinity ($\Lambda =0$) or the cosmological horizon ($\Lambda > 0$).
The potential (\ref{potential}) approaches zero as $r_{*}$ goes to
$\pm\infty$.
Therefore the asymptotic solution to Eq.~(\ref{Schroedinger}) is a linear
combination of $e^{+i\omega r_{*}}$ and
$e^{-i\omega r_{*}}$.
With the time dependence of the solution of the form $e^{i\omega t}$ ,
the solutions $e^{+i\omega r_{*}}$ and $e^{-i\omega r_{*}}$
represent an ingoing wave and an outgoing wave, respectively.
At infinity or the cosmological horizon we admit an outgoing wave, and at
the
event horizon an ingoing wave.
This is because we assume no external source of perturbation of the
membrane
during its evolution.
Hence we set the boundary conditions as follows
\begin{equation}
\chi \longrightarrow \left\{
\begin{array}{@{\,}ll}
C_{out}\exp (-i\omega r_{*} ), & \mbox{($r_{*}\to +\infty$)} \\
C_{in}\exp (+i\omega r_{*} ), & \mbox{($r_{*}\to -\infty$)}
\end{array}
\right. \label{bc}
\end{equation}
with $C_{in/out}$ constant.
This form of boundary conditions appears in the standard analysis of
a quasi-normal mode of a black hole~\cite{Cha}.

With the time dependence $e^{i\omega t}$, $Im(\omega )<0$ corresponds to
unstable modes. Then from the boundary conditions (\ref{bc}), an unstable
solution to Eq.~(\ref{Schroedinger}) decays as $e^{-|Im(\omega )| r_{*}}$
when $ r_{*}$ goes to $+\infty$ and as $e^{|Im(\omega )| r_{*}}$ when $
r_{*}$
goes to $-\infty$. Multiplying Eq.~(\ref{Schroedinger}) by $\bar{\chi}$
(upper bar means the complex conjugate) and integrating by parts, we obtain
\begin{equation}
\int_{-\infty}^{+\infty}\left( 
{\left|\frac{d\chi}{d r_{*}}\right| }^{2}
+V{\left|\chi\right| }^{2}\right) d r_{*}
+{\left[ -\bar{\chi}\frac{d\chi}{d r_{*}}\right] }_{-\infty}^{+\infty}
={\omega}^{2}\int_{-\infty}^{+\infty}{\left|\chi\right| }^{2}d r_{*} .
\end{equation}
As long as we consuder unstable modes, the surface term on the l.h.s. 
vanishes and the integrals on both sides converge. 
Therefore we obtain
\begin{equation}
{\omega}^{2}=\int_{-\infty}^{+\infty}\left( 
{\left|\frac{d\chi}{d r_{*}}\right| }^{2}+V{\left|\chi\right| }^{2}
\right) d r_{*}\Bigg/ 
\int_{-\infty}^{+\infty}{\left|\chi\right| }^{2}d r_{*} .\label{omegasq}
\end{equation}
Since the r.h.s. of Eq.~(\ref{omegasq}) is real, $\omega$ is real or pure 
imaginary. 
Here we seek for unstable modes ($Im(\omega )<0$).
So we can set $\omega =i\sigma\;(\sigma <0)$. 
Then the boundary conditions (\ref{bc}) read
\begin{equation}
\chi \longrightarrow \left\{
\begin{array}{@{\,}ll}
C_{out}\exp (+\sigma r_{*} ), & \mbox{($r_{*}\to +\infty$)} \\
C_{in}\exp (-\sigma r_{*} ), & \mbox{($r_{*}\to -\infty$)} 
\end{array}
\right. .\label{bc2}
\end{equation}
We shall examine the eigenvalue problem of Eq.~(\ref{Schroedinger}) 
subject to the boundary conditions (\ref{bc2}).
 
\subsection{Algorithm}
The eigenvalue problem considered is reduced to a two point boundary value 
problem.
We define four dependent variables as
\begin{equation}
y_1 =\chi ,\; y_2 =\frac{d\chi}{d r_{*}},\; y_3 =\sigma ,\; y_4 =A,
\end{equation}
where all $y_{i}$'s are functions of $r_{*}$, and $\sigma$ and $A$ are 
constants.
$A$ is the ratio of $C_{in}$ to $C_{out}$ (or $C_{out}$ to $C_{in}$).
The evolution equations are
\begin{equation}
y_{1}' =y_2 ,\; y_{2}' =[V+{(y_{3})}^2 ] y_1 ,\; y_{3}' =0 ,\; y_{4}' =0,
\label{numevol}
\end{equation}
where the prime denotes the differentiation with respect to $r_{*}$.
We impose two sets of boundary conditions.
The set 1 ($y_{4}=C_{out}/C_{in}\equiv A_{1}$) is
\begin{equation}
y_{1} =\exp (-y_{3} {r_{*}}_{1} )\; , y_{2} = -y_{3} y_{1} \; ,
(\mbox{at $r_{*}={r_{*}}_{1}$})\label{exbc1}
\end{equation}
\begin{equation}
y_{1} =y_{4}\exp (y_{3} {r_{*}}_{2} )\; , y_{2} =y_{3} y_{1} \; 
(\mbox{at $r_{*}={r_{*}}_{2}$}) .\label{exbc2}
\end{equation}
The set 2 ($y_{4}=C_{in}/C_{out}\equiv A_{2}$) is
\begin{equation}
y_{1} =y_{4}\exp (-y_{3} {r_{*}}_{1} )\; , y_{2} = -y_{3} y_{1} \; ,
(\mbox{at $r_{*}={r_{*}}_{1}$})\label{exbc3}
\end{equation}
\begin{equation}
y_{1} =\exp (y_{3} {r_{*}}_{2} )\; , y_{2} =y_{3} y_{1} \; 
(\mbox{at $r_{*}={r_{*}}_{2}$}) .\label{exbc4}
\end{equation}
The condition $A_{1}\times A_{2}=1$ is used for checking the 
reliability of calculations.
Note that the boundary conditions (\ref{exbc1}-\ref{exbc4}) leave 
$y_{3},y_{4}$ arbitrary.
In general an arbitrary choice of $y_{3},y_{4}$ at one boundary and the 
subsequent integration of the dependent variables do not ensure that the 
boundary conditions at the other boundary are satisfied. 
We solve this problem by the shooting method.
We also perform the calculation by the relaxation method to confirm the 
results of the shooting method.
 
\section{RESULTS}\label{results}
\subsection{Schwarzschild case}
For pure Reissner-Nordstr{\o}m cases the potential (\ref{potential}) is 
positive semi-definite for $m\ge 1$:
\begin{equation}
V(r)= \left( 1-\frac{2}{r}+\frac{e^2}{r^2} \right) 
\left[\left( m^2 -\frac{1}{4}\right)\frac{1}{r^2} +\frac{3}{2 r^3} 
-\frac{9 e^2}{4 r^4} \right] \ge 0 
\hspace{2em}\mbox{($r\ge r_+ $)} ,
\end{equation}
where $r_{+}=1+\sqrt{1-e^2}$ is the radius of the outer event horizon.
The equality is satisfied only if $m=e=1$.
Therefore, by the argument of the energy integral~\cite{Cha}, there is no
unstable mode subject to the boundary conditions (\ref{bc2}) when $m\ge 1$.

For the Schwarzschild case with $m=0$ the radial function
$R(r)=\chi (r)/\sqrt{r}$ satisfies
\begin{equation}
\left( 1-\frac{2}{r}\right)\frac{d^{2}R}{d r^{2}}+\frac{1}{r}\frac{dR}{dr}-
\frac{{\sigma}^{2}}{1-2/r}R=0 .
\end{equation}
In terms of a new variable $\zeta =\ln (r-2)$ this equation becomes
\begin{equation}
-\frac{d^{2}R}{d{\zeta}^{2}}+{\sigma}^{2}r^{2}R=0 .
\end{equation}
The solution to this equation is regarded as the zero energy eigenfunction
for
the potential ${\sigma}^{2}r^{2}$. This potential is positive definite since
$\sigma$ is real and negative.
Therefore we have no solution which decays when $\zeta$
approaches $\pm\infty$ (the event horizon and infinity) as required by the
boundary conditions (\ref{bc2})
\begin{equation}
R(r)=\frac{\chi (r)}{\sqrt{r}}\longrightarrow \left\{
\begin{array}{@{\,}ll}
e^{\sigma r}/\sqrt{r}, & \mbox{($r\to\infty\;(\zeta\to\ +\infty )$)} \\
{(r-2)}^{-2\sigma}/\sqrt{2}, & \mbox{($r\to r_{+}\;(\zeta\to -\infty )$)}
\end{array}
\right. .
\end{equation}
Hence there exists no unstable mode for the Schwarzschild case.

\subsection{RNdS case}
Before describing the numerical results, we comment on the upper bound of $m$ for which unstable modes may exisit in the Reissner-Nordstr{\o}m-de Sitter case.
The potential (\ref{potential}) takes the form
\begin{equation}
V(r)= \left\{
\begin{array}{@{\,}ll}
f F(r)/(4 r^{3}) & (e=0)\\
f G(r)/(4 r^{4}) & (e\ne 0)
\end{array}
\right. , \label{}
\end{equation}
where $f$ is defined by Eq.~(\ref{f}),
\begin{equation}
F(r)=-5\Lambda r^{3}+(4 m^{2}-1) r+6 
\label{F}
\end{equation}
and
\begin{equation}
G(r)=-5\Lambda r^{4}+(4 m^{2}-1) r^{2} +6 r -9 e^{2} .
\label{G}
\end{equation}

Firstly we consider $e=0$ case.
$F(r)$ is a monotonously decreasing function of $r$ and has the only one root 
for $r\ge 0$.
Therefore, by the arguement of the energy integral~\cite{Cha}, the sufficient condition for the stability of perturbation is 
\begin{equation}
F(r_{c})\ge 0,
\label{stconde0}
\end{equation}
where $r_{c}$ is the radius of the cosmological horizon.
Eq.~(\ref{stconde0}) reads
\begin{equation}
m^{2}\ge \frac{r_{c}-6+5\Lambda {r_{c}}^{3}}{4 r_{c}}=4-\frac{9}{r_{c}},
\label{stcond2e0}
\end{equation}
where $f(r_{c})=0$ is used.
When $e=0$, $r_{c}\ge 3$. 
Therefore Eq.~(\ref{stcond2e0}) shows that at least for $m\ge 2$ the 
potential~(\ref{potential}) is positive definite and that a membrane is 
stable against such perturbation.

Next we consider $e\ne 0$ case.
For $r\ge 0$, $G(r)$ either stays negative, or becomes positive for an interval
$(r_{1},r_{2})$ but otherwise negative.
Then the sufficient condition for the stability of perturbation is
\begin{equation}
G(r_{+})\ge 0\;\mbox{and}\;G(r_{c})\ge 0 ,
\label{stcondene0}
\end{equation}
which excludes the possibility that $G(r)$ stays negative, and ensures 
$r_{1}\le r_{+}\le r_{c}\le r_{2}$.
When $e^{2}$ increases, $r_{+}$ and $r_{2}$ decrease and $r_{1}$ and $r_{c}$ 
increase.
Hence it is sufficient to consider $e^{2}=1$ case.
In this case Eq.~(\ref{stcondene0}) reads
\begin{equation}
m^{2}\ge \max\left\{\frac{16 r_{c}-3 r_{c}+24}{4 {r_{c}}^{2}}, 
\frac{16 r_{+}-3 r_{+}+24}{4 {r_{+}}^{2}}\right\}=
4+\max\left\{-\frac{3}{4 r_{c}}+\frac{6}{{r_{c}}^{2}}, 
-\frac{3}{4 r_{+}}+\frac{6}{{r_{+}}^{2}}\right\} ,
\label{stcond2ene0}
\end{equation}
where again $f(r_{+})=f(r_{c})=0$ are used.
When $\Lambda >0$, $r_{+}>1+\sqrt{1-e^{2}}=1\;(\mbox{now}\; e^{2}=1)$.
For $r>1$
\begin{equation}
-\frac{3}{128}\le -\frac{3}{4 r}+\frac{6}{r^{2}}< 5.25.
\end{equation}
Therefore Eq.~(\ref{stcond2ene0}) shows that at least for $m\ge 4$ there 
exists no unstable mode.

\begin{figure}
\centerline{\epsfxsize 8cm \epsfbox{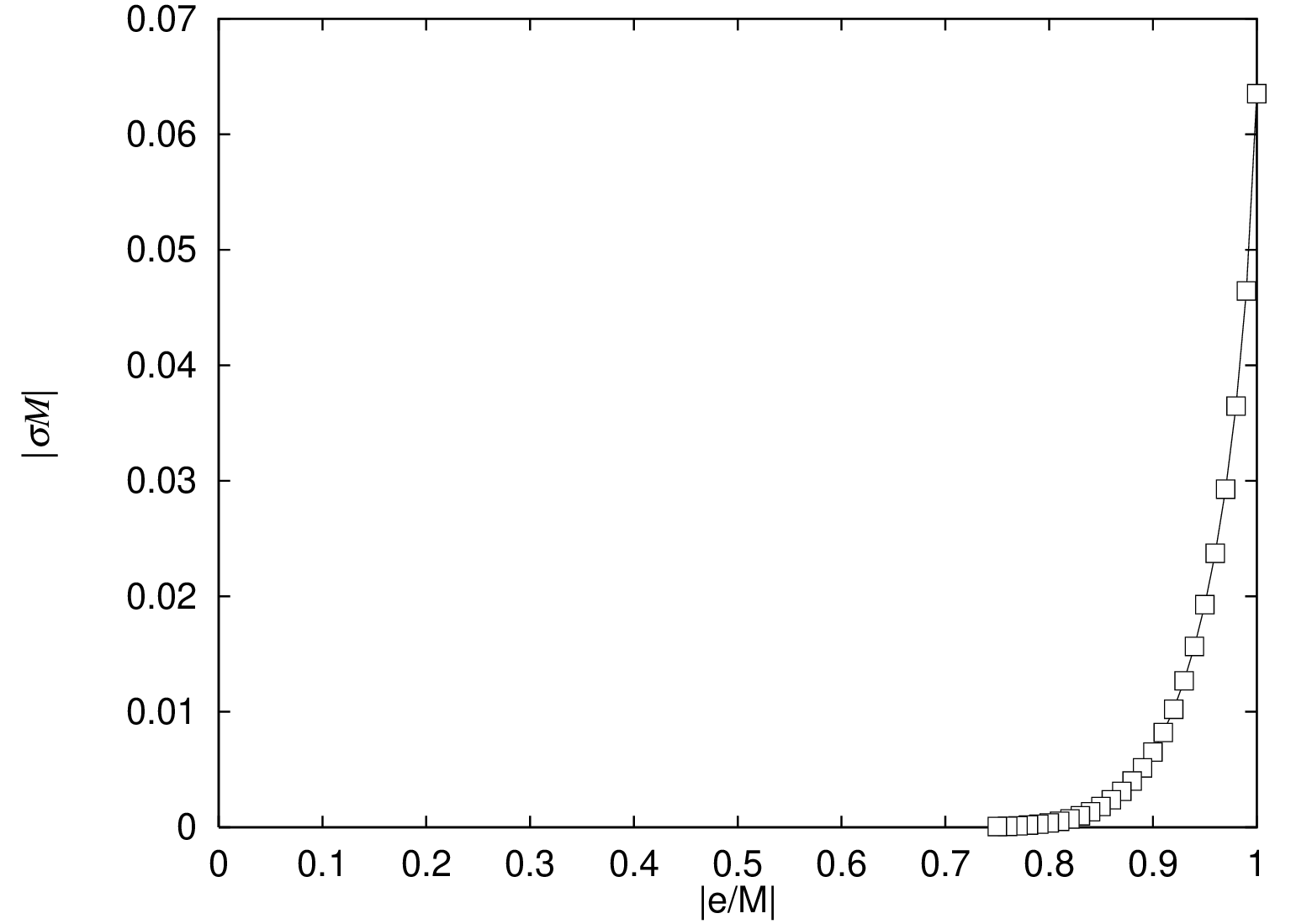}\epsfxsize 8cm
\epsfbox{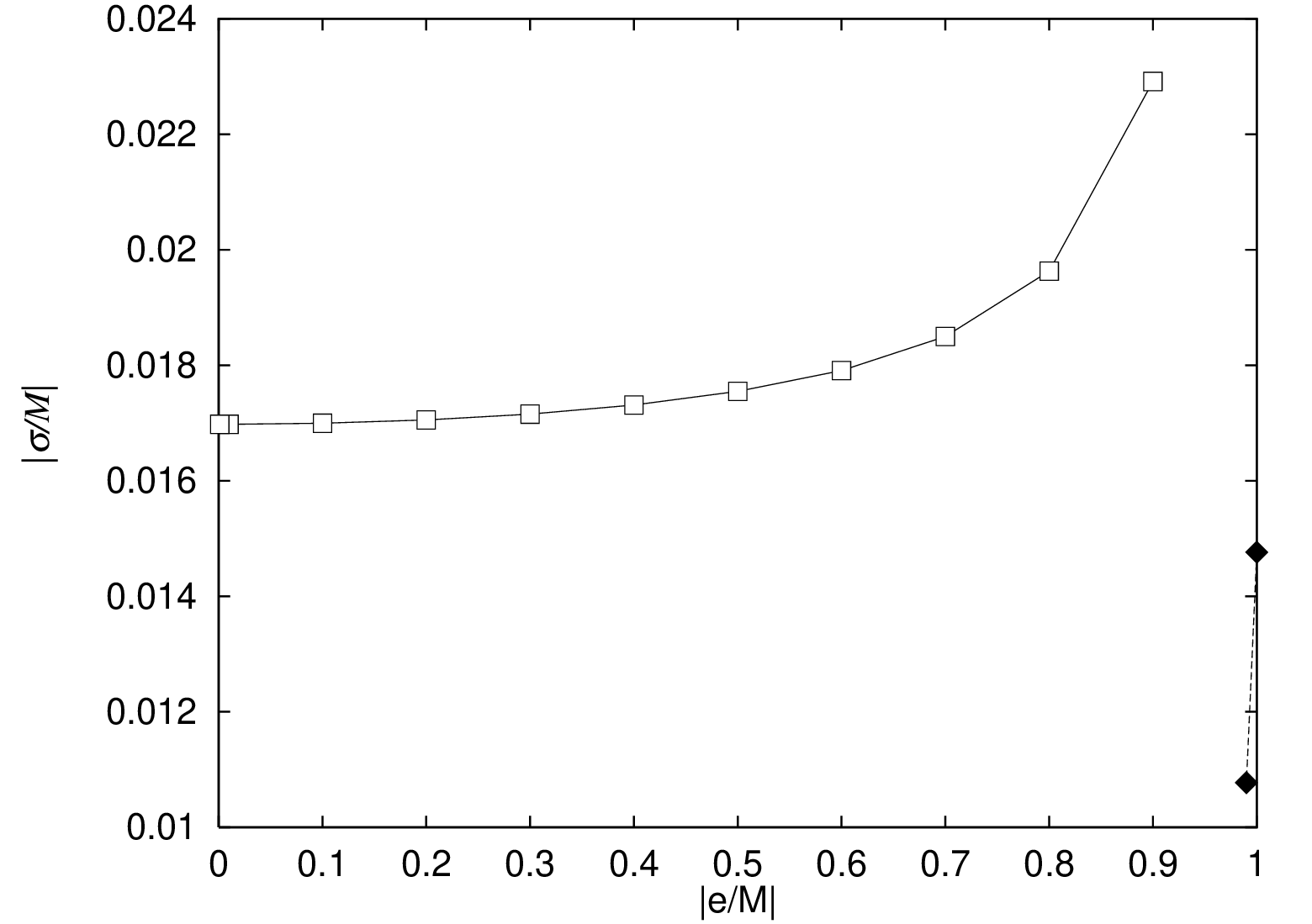}}
\centerline{\epsfxsize 8cm \epsfbox{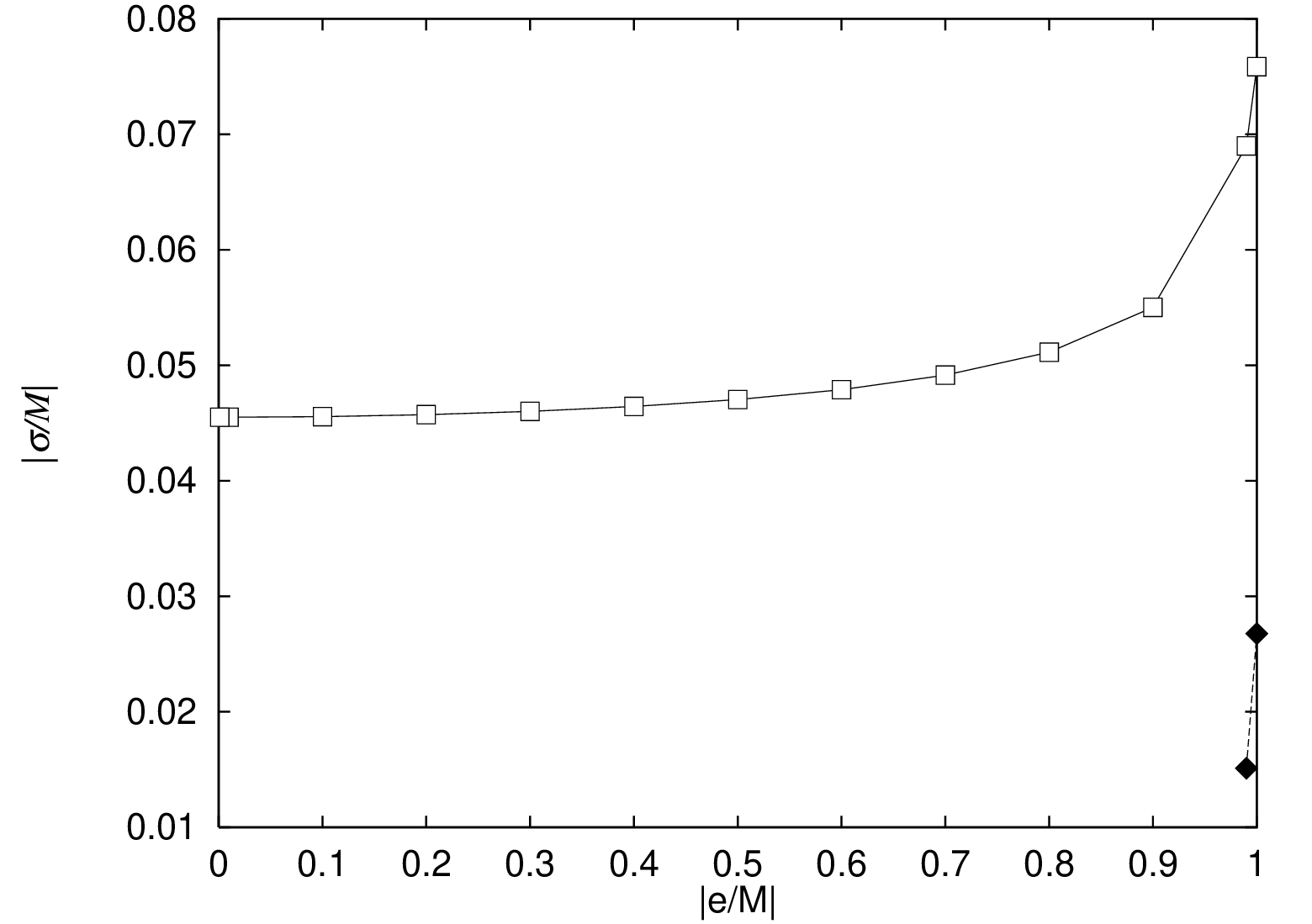}\epsfxsize 8cm
\epsfbox{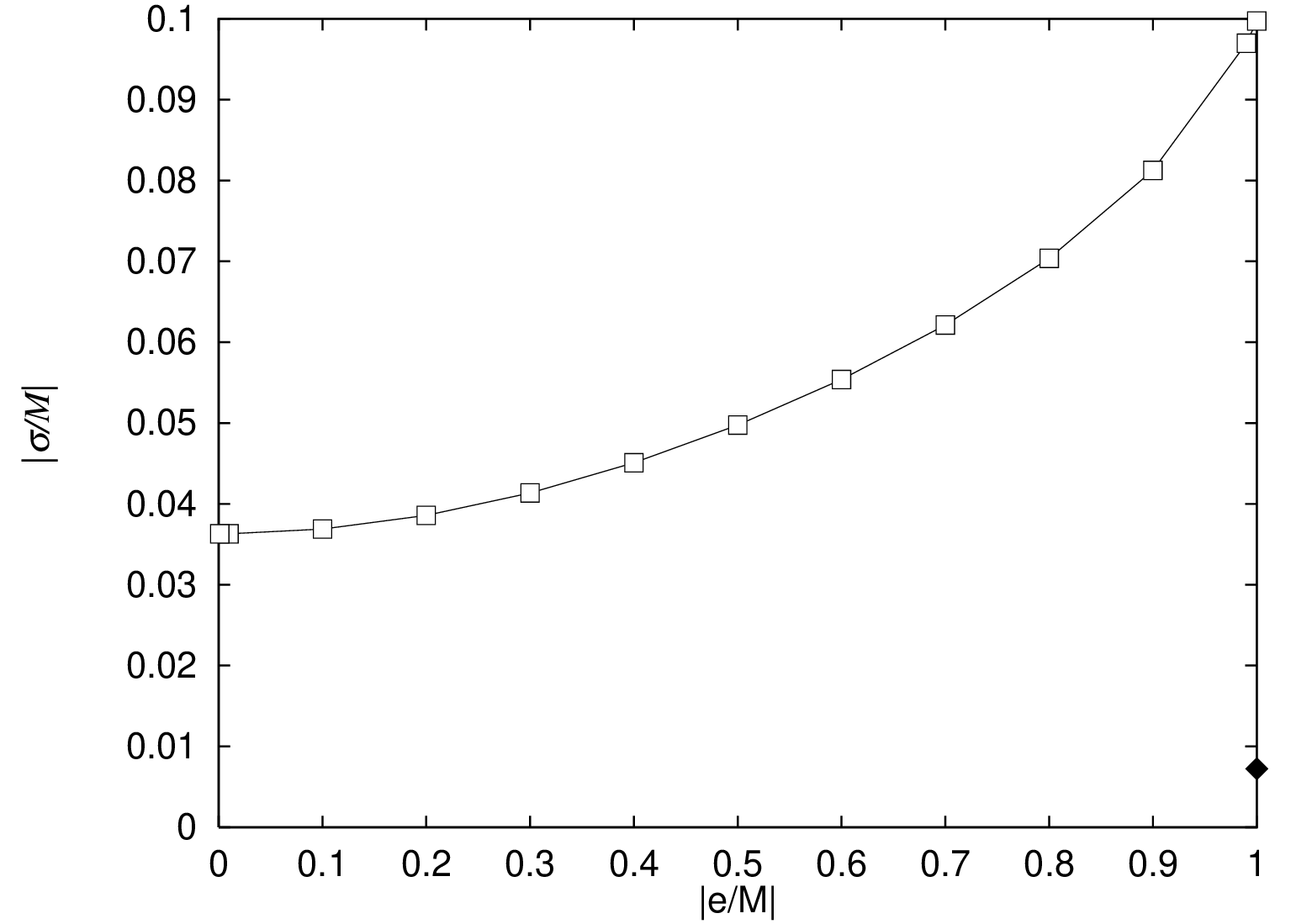}}
\caption{The plots of the eigenfrequency, $|\sigma M|$, versus the specific
charge, $|e|/M$, of the black hole for the $m=0$ cases. The top left panel
shows the plot
for the $\Lambda M^{2}=0$ case, the top right for the $\Lambda
M^{2}=1/1000$, the bottom
 left for the $\Lambda M^{2}=1/100$ and the bottom right for the $\Lambda
M^{2}=1/10$.
We obtained two sequences of the eigenvalues except for the $\Lambda
M^{2}=0$ case.}\label{plots}
\end{figure}

The condition that the Reissner-Nordstr{\o}m-de Sitter spacetime has a
static
region is $\Lambda <{\Lambda}_{max}$, where
\begin{equation}
{\Lambda}_{max}(e^2)=\frac{3(1+\sqrt{9-8 e^2})}
{(18-4 e^2)(3+\sqrt{9-8 e^2})-24 e^2} .
\end{equation}
This is a monotonously increasing function of $e^2$ and has the minimum
value
${\Lambda}_{max}(0)=1/9$.
So we numerically investigate the cases $\Lambda M^{2}=0, 1/1000, 1/100,
1/10$.
We found unstable modes for various parameter sets $(\Lambda ,|e|)$.
The eigenvalues of $m=0$ modes are shown in Fig. \ref{plots}.
The results of the shooting method and the relaxation method agree very
well,
the error being within 5\%.

Fig. \ref{plots} shows that when $|e|$ becomes larger, we have a
larger $|\sigma |$ and hence a higher growth rate of the perturbations.
Therefore the charge of the black hole indeed destabilizes the
membrane at the equatorial plane.
Since the membrane is electrically neutral, this instability should be
understood as the curvature effect of the charge of the black hole.

As to the effect of the cosmological constant $\Lambda$, we find no simple
tendency: for a given $|e|$, a larger $\Lambda$ does not necessarily lead to
a
larger $|\sigma |$.
However with a non-vanishing cosmological constant the magnitude of the
eigenvalue
is typically $10^{-2}$ for $0.001\le |e|/M \le 1$ in contrast to
the RN cases, in which the magnitude of the eigenvalue almost exponentially
decreases with decreasing specific charge and is far below $10^{-2}$ for
$|e|/M \lesssim 0.80$.
Therefore we can at least say that the cosmological constant also
destabilizes
 the membranes.
This expectation is supported by the consideration of the de Sitter
background case (see Appendix \ref{desitter}).

\begin{figure}
\centerline{\epsfbox{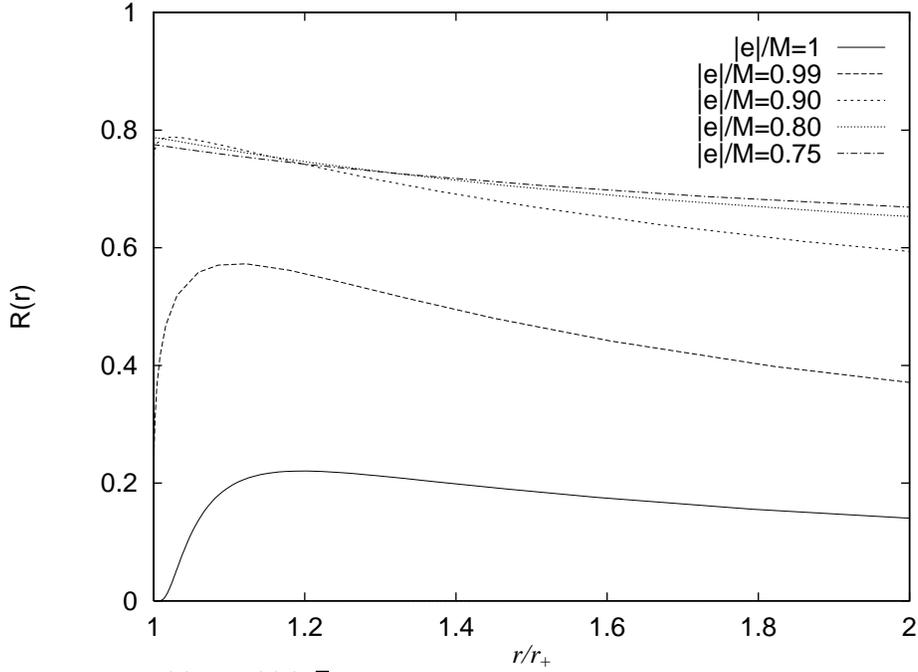}}
\caption[dummy]{The radial profiles $R(r)= \chi (r)/\sqrt{r}$
of the $m=0$ mode eigenfunctions for the pure Reissner-Nordstr{\o}m cases
($|e|/M=1,0.99,0.90,0.80,0.75$).
The abscissa is normalized in the unit of the horizon radii $r_{+}$. }
\label{profiles}
\end{figure}

We plotted some of the eigenfunctions for $\Lambda =0$ cases in
Fig. \ref{profiles}.
Some features of the profiles are listed in Table \ref{features}.
When we decrease the value of the specific charge of the black hole,
the peak of the corresponding profile gets higher and approaches the
event horizon.

\begin{figure}
\centerline{\epsfxsize 13cm \epsfbox{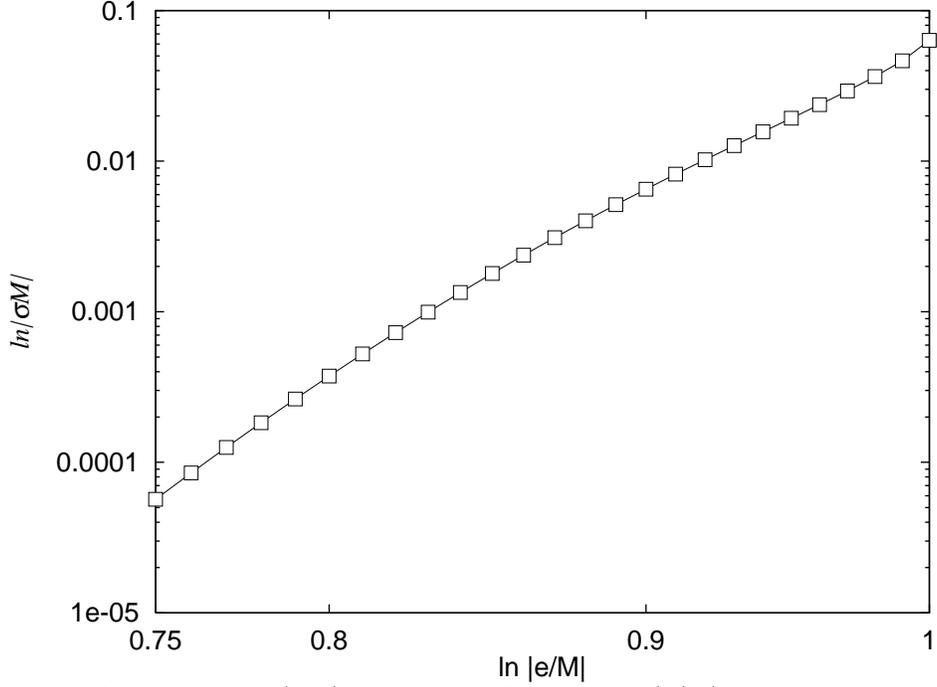}}
\caption{The plots of the eigenfrequency, $\ln |\sigma M|$, versus the
specific charge
, $\ln |e/M|$, of the black hole for the $\Lambda M^{2}=0,m=0$ cases. We
find an
approximate linear relation between $\ln |\sigma M|$ and $\ln |e/M|$.}
\label{loglog}
\end{figure}

As noted above we found that the charge of the black hole destabilizes the
membrane in general.
However, in the present analysis, we could not show the existence of
unstable
modes for small $|e|$ ($<0.75$) when $\Lambda =0$ due to the difficulty in
the numerical calculation as follows.
As $|e|$ becomes smaller, $|\sigma |$ decreases approximately in powers of
$|e|$
for $0.75\le |e| \le 1$ (See Fig. \ref{loglog}).
Judging by the extrapolation of this relation to the regime $|e|\le 0.75$,
$|\sigma |$ will be infinitesimal when $|e|\sim 0$.
In order for the asymptotic forms $e^{\pm i\omega r_{*}}$ of $\chi$ to be
justified, $|V(r)|$ must be much smaller than
$|{\omega }^{2}|=|{\sigma }^{2}|$
at the boundaries.
Then the coefficients in the evolution equations (\ref{numevol}) get
extremely
small and we suffer from underflows of numerical calculations.
By the same numerical difficulty, we leave open the possibility that unstable 
modes with $m\ge 1$ exist in the presence of the positive cosmological 
constant.

\section{DISCUSSION}\label{discussion}

In this paper we numerically studied the stability of a Nambu-Goto membrane
at the equatorial plane of the Reissner-Nordstr{\o}m-de Sitter spacetime
and found that in general such a membrane is unstable when the black hole is
charged.

A membrane is a two-dimensional extended object in the three-space.
So when it moves towards a black hole, it is expected to be inevitably
captured by the black hole.
A series of configurations found by Christensen et al. seem to represent
such
situations.
Among these configurations a membrane lying in the equatorial plane of the
black hole spacetime has the highest symmetry and is most likely to be the
final state of the scattering and capturing process.
On seeing our result, however, we expect that it is not the case.
In particular, Fig. \ref{profiles} allows us to imagine that the membrane 
moves away from the equatorial plane.
Then what eventually happens to that membrane?
Causality prohibits the membrane from escaping from the event horizon.
By analogy with the scattering process of a cosmic string off a black
hole~\cite{DF1,DF2} the membrane might experience large deformation and the
topology of the membrane might change.
Here are some speculations on the fate of the membrane; it may settle down
to
some other configuration than the equatorial plane, or it may break up into
two parts with one swallowed by the black hole and the other escaping to
infinity.
We need to perform a full dynamical computation to resolve this issue.
However that is beyond the scope of this paper.

We ignored the gravitational effect of a domain wall whereas a gravitating
domain wall is known to make a repulsive gravitational field~\cite{Ip},
which
is opposite to the strong attractive gravity of a black hole.
It is quite intriguing to find out the consequences of the competition of
these opposite forces.

\section*{ACKNOWLEDGEMENTS}
We would like to thank Profs. H. Sato, T. Nakamura, H. Kodama and T. Chiba 
for many useful comments.
We also thank Y. Morisawa and R. Yamazaki for fruitful discussion.
We are grateful to Dr. T. Harada for the discussion, especially, on the
boundary conditions.
Finally we thank the referee for constructive comments.
A.I. and D.I. were supported by the JSPS.
This work was supported in part by the Grant-in-Aid for Scientific Research
Fund (D.I., No. 4318).

\appendix
\section{de Sitter and Minkowski background cases}\label{desitter}
To understand the destabilizing effect of the positive cosmological
constant,
we compare the de Sitter and Minkowski cases.
In these cases Eq.~(\ref{Schroedinger}) is rewritten in terms of a new
function
$P(r_{*})=\chi (r_{*})/\sqrt{r_{*}}$ as
\begin{equation}
-\frac{d^{2}P}{d r_{*}^{2}}-\frac{1}{r_{*}}\frac{dP}{d r_{*}}+\left(
\frac{1}{4 r_{*}^{2}}-\frac{1}{4 r^{2}}-\frac{7}{6}\Lambda +{\sigma}^{2}
+\frac{5}{12}{\Lambda}^{2}r^{2}\right) P=0 ,
\label{Besseleq}
\end{equation}
where $m$ is set to zero for simplicity.
The reason we use $P(r_{*})$ is that in the de Sitter and Minkowski cases
the
potential (\ref{potential}) for $\chi (r_{*})$ diverges at the center $r=0$
and that numerical calculations become unstable.

In the Minkowski case $r_{*}=r$ and Eq.~(\ref{Besseleq}) becomes Bessel's
differential equation of order $0$
\begin{equation}
\frac{d^{2}P}{d{(\sigma r)}^{2}}+\frac{1}{\sigma r}\frac{dP}{d(\sigma
r)}-P=0 ,
\label{Mink}
\end{equation}
the solution of which is a linear combination of the modified Bessel
functions
$I_{0}(\sigma r)$ and $K_{0}(\sigma r)$.
However $I_{0}(\sigma r)$ and $K_{0}(\sigma r)$ diverge at $r\to\infty$ and
$r=0$, respectively and do not satisfy the regularity conditions.
Hence membranes are stable in the Minkowski case.
(The stability for $m\ge 1$ cases is also verified.)

When the positive cosmological constant is present, the situation changes.
In the limit of $r\to 0$ $(r_{*}\to -\infty)$, Eq.~(\ref{Besseleq}) reduces
to
\begin{equation}
\frac{d^{2}P}{d r_{*}^{2}}+\frac{1}{r_{*}}\frac{dP}{d r_{*}}+\left(
\frac{11}{9}\Lambda -{\sigma}^{2}\right) P=0 .
\label{dS}
\end{equation}
Unlike Eq.~(\ref{Mink}), the solution to Eq.~(\ref{dS}) differs depending on
the
sign of $(11\Lambda /9-{\sigma}^{2})$.
When $|\sigma |< \sqrt{11\Lambda /9}$, the solution is a
linear combination of the Bessel function $J_{0}(\alpha r_{*})$ and the
Neumann function $N_{0}(\alpha r_{*})$, where $\alpha\equiv
\sqrt{|11\Lambda /9-{\sigma}^{2}|}$.
When $|\sigma |> \sqrt{11\Lambda /9}$, the solution is a
linear combination of the modified Bessel functions $I_{0}(\alpha r_{*})$
and
$K_{0}(\alpha r_{*})$.
The regularity condition at the center excludes $N_{0}$ and $K_{0}$ as the
asymptotic solutions.

When $|\sigma |< \sqrt{11\Lambda /9}$, the asymptotic behavior of $P$ near
the center changes from the Minkowski case and we expect the emergence of
unstable modes since $J_{0}$ is a bounded function.
In fact we found unstable modes by numerical calculations.
The method is similar to the one for the RNdS cases. The asymptotic form of
$P$ near the cosmological horizon $r_{*}\to\infty$ is determined to be
$e^{\sigma r_{*}}/\sqrt{r_{*}}$ by the boundary condition similar to
Eq.~(\ref{bc2}).
The eigenvalues are $-0.18257, -0.05773, -0.01826$ for $\Lambda =1/10,
1/100, 1/1000$, respectively.
Though we performed numerical calculations just for three values of the
cosmological constant, unstable modes are expected as long as the positive
cosmological constant is present.

\section{CODE CHECK}
We calculated the energy spectra for a one-dimensional square-well potential
to check the reliability of our numerical code.

We consider an eigenvalue problem
\begin{equation}
-\frac{d^{2}}{d{x}^{2}}\phi +V(x)\phi =-{\sigma}^{2}\phi
\label{testeq}
\end{equation}
for a square-well potential
\begin{equation}
V(x)=\left\{
\begin{array}{@{\,}ll}
-1 & \mbox{($|x|\le w$, $w>0$)} \\
0  & \mbox{(otherwise)}
\end{array}
\right. .
\label{wellpot}
\end{equation}
The problem reduces to a two point boundary value problem at the
boundaries $x=x_{1},x_{2}$ ($x_{1}<-w, x_{2}>w$).
The dependent variables are
\begin{equation}
y_{1}=\phi ,\; y_{2}=\frac{d\phi}{dx},\; y_{3}=\sigma ,\;
y_{4}=A,
\label{wellevo}
\end{equation}
where $\sigma$ and $A$ are constants.
The evolution equations and the boundary conditions are similar to Eq.
(\ref{numevol}) and Eqs. (\ref{exbc1},\ref{exbc2}) or Eqs.
(\ref{exbc3},\ref{exbc4}), respectively.

We can analytically show that the eigenvalues for the potential
(\ref{wellpot}) satisfy the equation
\begin{equation}
n\pi -2w s-2\arcsin s=0 \hspace{2em}(n=0,\pm 1,\pm 2,\cdots)
\label{wellexact}
\end{equation}
where $s=\sqrt{1-{\sigma}^{2}}$. The eigenvalues ${\sigma}_{wn}$ of
Eq.~(\ref{testeq}) are labeled by $w$ and $n$.
We found good agreements of eigenvalues ${\sigma}_{wn}$ obtained by solving
the two point boundary value problem, with those obtained by directly
solving Eq.~(\ref{wellexact}) (Table~\ref{testtable}).

\section{code check 2}
In this paper we present the results for which the shooting and relaxation 
methods agree within 5 \% error.
The agreement is, however, far better than 5\% in most cases.
The comparison of the two methods is summarized in Table~\ref{comp} for some 
cases.

\newpage

\begin{table}
\caption{Some features of the radial profiles in Fig.~2.}
\begin{center}
\begin{tabular} {|ccccc|}
charge $(|e|/M)$ & event horizon $(r_{+}/M)$ & peak location
$r_{\mbox{{\scriptsize peak}}}/M$ &
peak amplitude $R(r_{\mbox{{\scriptsize paek}}})$ & $r_{\mbox{peak}}/r_{+}$
\\ \hline
1    & 1     & 1.202 & 0.22 &  1.202 \\
0.99 & 1.141 & 1.281 & 0.57 &  1.123 \\
0.90 & 1.436 & 1.468 & 0.79 &  1.022 \\
0.80 & 8/5   & 1.603 & 0.79 &  1.002 \\
0.75 & 1.661 & 1.662 & 0.78 &  1.001 \\
\end{tabular}
\end{center}\label{features}
\end{table}

\begin{table}[t]
\caption{Code check. Comparison of numerically obtained eigenvalues for the
square-well potential (B2) with those obtained from Eq.~(B4).}
\begin{center}
\begin{tabular}{|c|c||ccc|}
 & analytic (from Eq.~(\ref{wellexact})) & numerical & A (see Eq.~(\ref{wellevo}))&
parity \\ \hline
${\sigma}_{11}$ & 0.673612 & 0.67361 & +1 & even \\
${\sigma}_{21}$ & 0.85723  & 0.85723 & +1 & even \\
${\sigma}_{22}$ & 0.319023 & 0.31902 & -1 & odd  \\
${\sigma}_{31}$ & 0.920798 & 0.92080 & +1 & even \\
${\sigma}_{32}$ & 0.650366 & 0.65037 & -1 & odd  \\
${\sigma}_{41}$ & 0.949724 & 0.94972 & +1 & even \\
${\sigma}_{42}$ & 0.785671 & 0.78566 & -1 & odd  \\
${\sigma}_{43}$ & 0.438306 & 0.43830 & +1 & even \\
${\sigma}_{51}$ & 0.965261 & 0.96527 & +1 & even \\
${\sigma}_{52}$ & 0.854684 & 0.85467 & -1 & odd  \\
${\sigma}_{53}$ & 0.641057 & 0.64105 & +1 & even \\
${\sigma}_{54}$ & 0.192693 & 0.19269 & -1 & odd  \\
\end{tabular}
\end{center}
\label{testtable}
\end{table}

\begin{table}
\caption{Comparison of the shooting and relaxation methods.
Some selected eigenvalues $\sigma M$ are listed.}
\begin{tabular}{|c|cc|}
Specific charge ($|e|/M$) & Shooting & Relaxation \\ \hline
$\Lambda M^2 =0$ & &  \\ \hline 
1    &           $ 6.351 \times 10^{-2}$ & $ 6.328 \times 10^{-2}$ \\
0.99 &           $ 4.644 \times 10^{-2}$ & $ 4.640 \times 10^{-2}$ \\
0.97 &           $ 2.928 \times 10^{-2}$ & $ 2.927 \times 10^{-2}$ \\
0.95 &           $ 1.928 \times 10^{-2}$ & $ 1.928 \times 10^{-2}$ \\
0.93 &           $ 1.268 \times 10^{-2}$ & $ 1.268 \times 10^{-2}$ \\
0.91 &           $ 8.196 \times 10^{-3}$  & $ 8.182 \times 10^{-3}$  \\
0.89 &           $ 5.139 \times 10^{-3}$  & $ 5.120 \times 10^{-3}$  \\
0.87 &           $ 3.104 \times 10^{-3}$  & $ 3.084 \times 10^{-3}$  \\
0.85 &           $ 1.797 \times 10^{-3}$  & $ 1.778 \times 10^{-3}$  \\
0.83 &           $ 9.939 \times 10^{-4}$ & $ 9.792 \times 10^{-4}$ \\
0.81 &           $ 5.242 \times 10^{-4}$ & $ 5.180 \times 10^{-4}$ \\
0.79 &           $ 2.632 \times 10^{-4}$ & $ 2.559 \times 10^{-4}$ \\
0.77 &           $ 1.255 \times 10^{-4}$ & $ 1.211 \times 10^{-4}$ \\
0.75 &           $ 5.676 \times 10^{-5}$ & $ 5.416 \times 10^{-5}$ \\ \hline
$\Lambda M^2 =0.1$ & & \\ \hline
1     &           $ 9.9727 \times 10^{-2}$ & $9.9764 \times 10^{-2}$ \\
1(2nd seq.) &     $ 7.2435 \times 10^{-3}$ & $7.3476 \times 10^{-3}$ \\
0.99 &            $ 9.6971 \times 10^{-2}$ & $9.7009 \times 10^{-2}$ \\
0.90 &            $ 8.1253 \times 10^{-2}$ & $8.1291 \times 10^{-2}$ \\
0.80 &            $ 7.0378 \times 10^{-2}$ & $7.0409 \times 10^{-2}$ \\
0.70 &            $ 6.2124 \times 10^{-2}$ & $6.2149 \times 10^{-2}$ \\
0.60 &            $ 5.5387 \times 10^{-2}$ & $5.5406 \times 10^{-2}$ \\
0.50 &            $ 4.9761 \times 10^{-2}$ & $4.9775 \times 10^{-2}$ \\
0.40 &            $ 4.5092 \times 10^{-2}$ & $4.5103 \times 10^{-2}$ \\
0.30 &            $ 4.1352 \times 10^{-2}$ & $4.1361 \times 10^{-2}$ \\
0.20 &            $ 3.8587 \times 10^{-2}$ & $3.8594 \times 10^{-2}$ \\
0.10 &            $ 3.6878 \times 10^{-2}$ & $3.6885 \times 10^{-2}$ \\
0.01 &            $ 3.6305 \times 10^{-2}$ & $3.6311 \times 10^{-2}$ \\
0.001 &           $ 3.6298 \times 10^{-2}$ & $3.6305 \times 10^{-2}$ \\
\end{tabular}
\label{comp}
\end{table}

\end{document}